\documentclass[useAMS,usenatbib]{mn2e}

\usepackage{graphicx} 
 
\newcommand{\vsini}{$v \sin i$}

\usepackage[usenames]{color}

\title[The magnetic field of HD\,154708]{The determination of the rotation period and magnetic field 
geometry of the strongly magnetic roAp star HD\,154708\thanks{Based on 
observations collected at ESO, Paranal, Chile (ESO programmes Nos.\ 075.D-0145, 
076.D-0169, 077.D-0150, and 079.D-0240).}} 
\author[S. Hubrig et al.]{S. Hubrig$^{1,2}$\thanks{E-mail: shubrig@aip.de}, 
        G. Mathys$^{1}$,
        D.~W. Kurtz$^{3}$, 
        M. Sch\"oller$^{4}$, 
        V.~G. Elkin$^{3}$,
        H.~F. Henrichs$^{5}$\\ 
$^{1}$European Southern Observatory, Casilla 19001, Santiago, Chile\\ 
$^{2}$Astrophysical Institute Potsdam, An der Sternwarte 16, D-14482 Potsdam, Germany\\
$^{3}$Centre for Astrophysics, University of Central Lancashire, Preston PR1 2HE, 
UK\\ 
$^{4}$European Southern Observatory, Karl-Schwarzschild-Str.\ 2, 85748 Garching 
bei M\"unchen, Germany\\
$^{5}$Astronomical Institute, University of Amsterdam, Kruislaan 403, 1098 SJ Amsterdam, Netherlands 
} 
\begin{document} 
 
\date{Accepted 2007 Enero 99. Received 2007 Enero 98} 
 
\pagerange{\pageref{firstpage}--\pageref{lastpage}} \pubyear{2007} 
 
\maketitle 
 
\label{firstpage} 
 
\begin{abstract} 
We obtained thirteen spectropolarimetric observations of the strongly magnetic 
rapidly oscillating Ap star HD\,154708 over three months with the multi-mode 
instrument FORS\,1, installed at the 8-m Kueyen telescope of the VLT. These 
observations have been used for the determination of the rotation period of 
$ P=5.3666 \pm 0.0007$\,d. Using stellar fundamental parameters and the longitudinal 
magnetic field phase curve, we briefly discuss the magnetic field geometry. The 
star is observed nearly pole-on and the magnetic field geometry can be described 
by a centred dipole with a surface polar magnetic field strength $B_{\rm d}$ between
26.1 and 28.8\,kG and an inclination of the magnetic axis to the rotation axis 
in the range 22.5$^{\circ}$ to 35.5$^{\circ}$. 
\end{abstract} 
 
 
\section{Introduction} 
\label{sect:intro} 
 

The magnetic A and F stars have global magnetic fields with surface field 
strengths typically of a few kG to tens of kG. A subset of them, the rapidly 
oscillating Ap (roAp) stars, pulsate in high radial overtone magnetically 
distorted dipole and quadrupole modes with periods in the range $5.65 - 21$\,min 
(see, e.g., Table~1 of \citealt{Kurtz2006}). These stars thus offer the 
opportunity to observe the interaction of p\,modes with strong magnetic fields as 
can be done for no other star but the sun. The kG-strength magnetic fields in 
sunspots dissipate energy in solar p-modes and shift their phases significantly, 
as is clearly seen in both observational and theoretical local helioseismology. 
While it is not possible to perform local asteroseismology in the same detail as 
for the sun -- for obvious reasons of angular resolution -- some unique properties 
of the roAp stars have the potential to allow 3D mapping of their pulsation modes, 
magnetic field geometries and abundance distributions within the observable layers 
of their atmospheres, hence allow a detailed study of the interaction of the 
pulsations with the magnetic fields. 
 
The roAp stars with very strong magnetic fields are of special interest because 
the effect of the magnetic field on the structure of their atmospheres can be 
studied with the greatest detail and accuracy. The cool roAp star HD\,154708 
possesses the strongest magnetic field among the roAp stars, with a mean magnetic 
field modulus $\left<B\right>=24.5\pm1.0$\,kG  \citep{Hubrig2005}, 
which is about a factor of three greater than that of HD\,166473,
$\left<B\right>\sim5.5$--9.0\,kG \citep{Mathys2007}, the roAp star
with the second-strongest magnetic field.
For all of the roAp stars 
the pulsations observed in radial velocity variations of lines of the rare earth 
elements arise at atmospheric heights where the magnetic pressure exceeds the gas 
pressure and the Alfv\'en velocity is greater than the acoustic velocity, so the 
pulsations are primarily magnetic with an acoustic component. There have been many 
theoretical investigations of the roAp stars discussing, e.g., mode conversion and 
reflection, contributions of horizontal and vertical motions to the observed 
pulsational radial velocities, and mode excitation and suppression in these stars 
(see, e.g., \citealt{SousaCunha2008} and \citealt{Saio2005}, and references therein). 

All of the theoretical models are for atmospheres with magnetic field strengths 
less than 10\,kG. Thus the discovery of pulsation in HD\,154708 in the presence of 
its 24.5-kG field charts new territory. With its field strength more than 15 times 
the 1.5\,kG typically observed in sunspots, HD\,154708 by far shows the most 
extreme observable case of pulsation in the presence of a strong magnetic field \citep{Kurtz2006}.
We thus wish to know as much about it as possible, and for obliquely rotating and 
pulsating Ap stars it is imperative to determine their rotation periods and 
magnetic variations with rotation to extract important geometrical information. 

The atmospheric parameters of HD\,154708, $T_{\rm eff}= 6800$\,K and $\log g= 
4.11$, have been determined using the B2$-$G temperature calibration of Geneva 
photometry (\citealt{HauckNorth1993}, Eqn.~2) and the Hipparcos parallax \citep{Hubrig2005}; 
Str\"omgren photometry \citep{Napiwotzki1993} suggests a hotter temperature, 
$T_{\rm eff}= 7500$\,K. The magnetic field in this star was discovered during our 
systematic study of magnetic fields in chemically peculiar Ap and Bp stars with 
the multi-mode instrument FORS\,1 installed at the 8-m Kueyen telescope 
\citep{Hubrig2005}. Among the whole class of Ap and Bp stars, the only star known 
at that time to have a stronger dipolar magnetic field than HD\,154708 was the 
much hotter ($T_{\rm  eff} \approx $14\,500\,K; \citealt{Leckrone1974}) Bp Si star 
HD\,215441 (Babcock's star) with a mean field modulus of $\left<B\right>=34$\,kG 
\citep{Babcock1960}. Now \citet{Freyhammer2008} have found a field of 
$\left<B\right>=30.29\pm0.08$\,kG in HD\,75049, making this the second-strongest 
dipolar field known among the upper main sequence chemically peculiar stars.

The presence of pulsations with a period of 8\,min was discovered by 
\citet{Kurtz2006} using 2.5\,h of high time resolution UVES (Ultraviolet and 
Visual Echelle Spectrograph) spectra. They found that the radial velocity 
amplitudes in the rare earth element lines of Nd\,\textsc{ii}, Nd\,\textsc{iii}, 
and Pr\,\textsc{iii} were unusually low, suggesting that roAp stars with stronger 
magnetic fields have lower pulsation amplitudes. Further study of the pulsation of 
HD\,154708 over its now-known 5.3666-d rotation cycle will illuminate this suggestion, 
as would theoretical studies of models with field strengths up to 30\,kG. 
 
A recent abundance analysis of HD\,154708 was based on a high spectral 
resolution UVES spectrum obtained on September 18, 2005 \citep{Nesvacil2008}.
The atmospheric chemical composition of this star 
was shown to be typical of cool Ap stars with a significant ionization 
disequilibrium for the first and second rare earth ions, which is commonly 
observed in the atmospheres of pulsating roAp stars \citep{Ryab2004}. The model 
atmosphere was calculated with the LLModels code \citep{Shulyak2004}. Light 
elements, as well as Ti, Fe, and Ni were found underabundant whereas Sc and Cr 
showed solar abundances. Heavy elements and rare earth elements were strongly 
overabundant. Nd\,\textsc{iii} is 0.86\,dex overabundant with respect to 
Nd\,\textsc{ii} and Pr\,\textsc{iii} was overabundant by 1.11\,dex compared to 
Pr\,\textsc{ii}. 

\citet{Cowley2007} studied the presence of heavy Ca isotopes in HD\,154708 using 
the infrared triplet lines of Ca\,\textsc{ii}. They measured isotopic shifts of 
+0.14\,\AA{}, +0.26\,\AA{}, and +0.29\,\AA{} for $\lambda$8498, 8542, and 8662, 
respectively. However, the maximum isotopic shifts between the isotopes $^{40}$Ca 
and $^{48}$Ca for the Ca\,\textsc{ii} triplet lines is of the order of 0.2\,\AA{}. 
Possibly these differences are caused by other effects than isotopic shifts -- 
most likely, blending with highly displaced Zeeman components. The magnetic field 
of HD\,154708 is so strong that a number of spectral lines are distorted beyond 
recognition by the partial Paschen-Back effect. 
 
The magnetic field geometry in HD\,154708 has not previously been studied. In this 
work we present for the first time a mean longitudinal magnetic field measurement 
series, obtained with the multi-mode instrument FORS\,1 installed at the 8-m 
Kueyen telescope. The data are used to obtain the rotation period and to put 
constraints on the magnetic field geometry. This is the first, necessary step for 
a more detailed study of this remarkable star.
 
\begin{table} 
\centering 
\caption{ 
Magnetic field measurements of HD\,154708 with FORS\,1. All quoted errors are 
1$\sigma$ uncertainties.
In the last three lines we list the earlier measurements published in
\citet{Hubrig2006}.}
\label{tab:hd154_magfield} 
\begin{tabular}{l r@{$\pm$}l r@{$\pm$}l} 
\hline 
\multicolumn{1}{c}{MJD} & 
\multicolumn{2}{c}{$\left< B_{\rm z}\right>_{\rm all}$} & 
\multicolumn{2}{c}{$\left< B_{\rm z}\right>_{\rm hydr}$}\\ 
& 
\multicolumn{2}{c}{[G]} & 
\multicolumn{2}{c}{[G]} \\ 
\hline 
\hline 
54209.3072 & 7941 & 15 & 7711 & 41 \\ 
54215.2792 & 7557 & 15 & 7301 & 49 \\ 
54223.1734 & 6765 & 15 & 6174 & 41 \\ 
54238.2840 & 6427 & 23 & 6062 & 65 \\ 
54247.1726 & 7726 & 20 & 7483 & 62 \\ 
54254.1219 & 6497 & 28 & 6103 & 80 \\ 
54258.2500 & 7462 & 20 & 7330 & 59 \\ 
54270.3097 & 6479 & 15 & 6169 & 46 \\ 
54279.1667 & 7884 & 15 & 7710 & 44 \\ 
54287.2260 & 6326 & 59 & 5718 & 169 \\ 
54297.3028 & 6450 & 28 & 6211 & 83 \\ 
54305.1536 & 7873 & 18 & 7492 & 49 \\ 
54307.0214 & 7032 & 18 & 6884 & 41 \\ 
\hline 
53120.376  & \multicolumn{2}{c}{} & 7530 & 54 \\
53487.302  & \multicolumn{2}{c}{} & 5764 & 25 \\
53519.344  & \multicolumn{2}{c}{} & 5819 & 52 \\
\hline 
\end{tabular} 
\end{table} 
 
\section{Observations} 
 
The observations were carried out in 2007 April -- July in service mode. Using the 
narrowest available slit width of 0\farcs4 the achieved spectral resolving power 
of the FORS\,1 spectra obtained with the GRISM 600B was about 2000. These 
observations made use of a new mosaic detector with blue optimized E2V chips, 
which was installed in FORS\,1 at the beginning of April 2007. It has a pixel size 
of 15\,$\mu$m (compared to 24\,$\mu$m for the previous Tektronix chip) and higher 
efficiency in the wavelength range below 6000\,\AA{}. With the new mosaic detector 
and the GRISM 600B, we are also able to cover a much larger spectral range, from 
3250 to 6215\,\AA{}. A detailed description of the assessment of the longitudinal 
magnetic field measurements using FORS\,1 is presented in our previous papers 
(e.g., \citealt{Hubrig2004a, Hubrig2004b}, and references therein). We repeat here 
the major steps of the magnetic field determination. The mean longitudinal 
magnetic field, $\left< B_{\rm z}\right>$, was derived using 
\begin{equation} 
\frac{V}{I} = -\frac{g_{\rm eff} e \lambda^2}{4\pi{}m_ec^2}\ \frac{1}{I}\ 
\frac{{\rm d}I}{{\rm d}\lambda} \left<B_{\rm z}\right>, 
\label{eqn:one} 
\end{equation} 
\noindent 
where $V$ is the Stokes parameter which measures the circular polarization, $I$ 
is the intensity in the unpolarized spectrum, $g_{\rm eff}$ is the effective 
Land\'e factor, $e$ is the electron charge, $\lambda$ is the wavelength, $m_e$ the 
electron mass, $c$ the speed of light, ${{\rm d}I/{\rm d}\lambda}$ is the 
derivative of Stokes $I$, and $\left<B_{\rm z}\right>$ is the mean longitudinal magnetic 
field. To minimize the cross-talk effect, we executed a sequence of spectropolarimetric observations
at different position angles of the retarder waveplate, +45$-$45, 
+45$-$45, +45$-$45, etc., and calculated the values $V/I$ using: 
 
\begin{equation} 
\frac{V}{I} = 
\frac{1}{2} \left\{ \left( \frac{f^{\rm o} - f^{\rm e}}{f^{\rm o} + f^{\rm e}} 
\right)_{\alpha=-45^{\circ}} 
- \left( \frac{f^{\rm o} - f^{\rm e}}{f^{\rm o} + f^{\rm e}} 
\right)_{\alpha=+45^{\circ}} \right\}, 
\label{eqn:two}   
\end{equation} 
 
\noindent 
where $\alpha$ denotes the position angle of the retarder waveplate and $f^{\rm 
o}$ and $f^{\rm e}$ are ordinary and extraordinary beams, respectively. Stokes $I$ 
values were obtained from the sum of the  ordinary and extraordinary beams. To 
derive $\left<B_{\rm z} \right>$, a least-squares technique was used to minimize the 
expression 
\begin{equation} 
\chi^2 = \sum_i \frac{(y_i - \left<B_{\rm z} \right> x_i - b)^2}{\sigma_i^2} 
\label{eqn:three}   
\end{equation} 
 
\noindent 
where, for each spectral point $i$, $y_i = (V/I)_i$, $x_i = -\frac{g_{\rm eff} e 
\lambda_i^2}{4\pi{}m_ec^2}\ (1/I\ \times\ {\rm d}I/{\rm d}\lambda)_i$, and $b$ is 
a constant term that, assuming that Eq.~\ref{eqn:one} is correct, approximates the 
fraction of instrumental polarization not removed after the application of 
Eq.~\ref{eqn:two} to the observations. In our calculations we assumed a Land\'e 
factor $g_{\rm eff}=1$ for hydrogen lines and $g_{\rm eff}=1.25$ for metal lines. 
The average Land\'e factor for metal lines is obtained using individual 
Land\'e factors for the observed spectral lines and is close to the value  $g_{\rm eff}=1.23$
published by \citet{Romanyuk1984}.
Furthermore, in our reduction procedure the spectral regions containing telluric 
lines have been excluded in the measurements of the magnetic field. The errors of 
the measurements of the polarization have been determined from photon 
statistics and have been converted to errors of field measurements. Longitudinal 
magnetic fields were measured in two ways: using only the absorption hydrogen 
Balmer lines, or using the entire spectrum including all available absorption 
lines. Due to the low resolution of the FORS1 polarimetric spectra and due to
the fact that the major contribution to the measured magnetic field
strength comes from the hydrogen
lines, the weak-field approximation can be reasonably adopted for
magnetic field strengths up to a few tens of kG.
The results of our magnetic field measurements are presented in 
Table~\ref{tab:hd154_magfield}. In the first column we provide the modified Julian 
dates at the middle of the exposures.
The measured mean 
longitudinal magnetic field $\left<B_{\mathrm z}\right>$ using the whole spectrum 
is presented in Col.~2. The measured mean longitudinal magnetic field 
$\left<B_{\mathrm z}\right>$ using all hydrogen lines is listed in Col.~3. All 
quoted errors are 1$\sigma$ uncertainties.
In the last three lines we list the earlier measurements published in
\citet{Hubrig2006}.
 
\section{Period determination}

\begin{figure} 
\centering 
\includegraphics[width=0.45\textwidth]{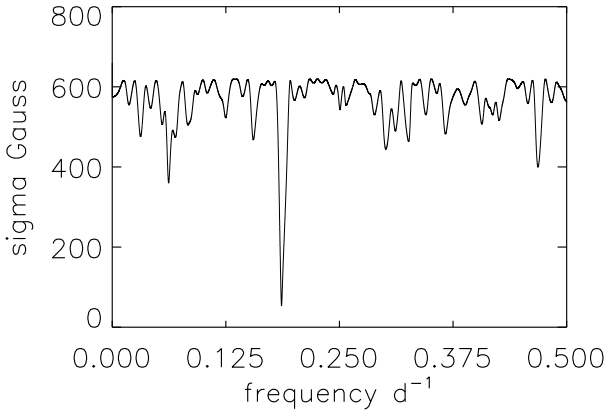} 
\caption{For the data in Col.~2 of Table~\ref{tab:hd154_magfield},
this plot shows the standard deviation of 
one measurement with respect to a sinusoidal fit for a range of frequencies. There 
is a clear best fit for a frequency of 0.1863\,d$^{-1}$, or $P = 5.367$\,d.
} 
\label{fig:lsps1} 
\end{figure}

\begin{figure}
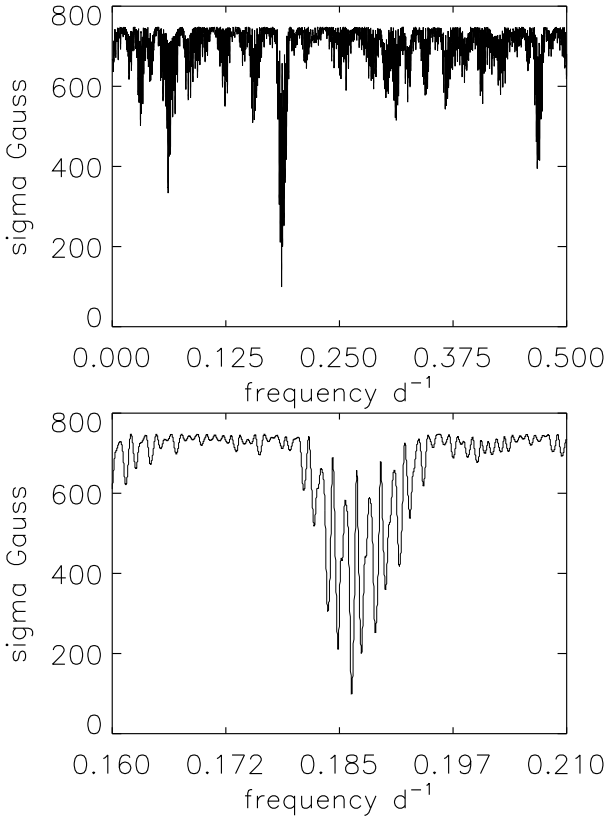
 
\centering 
\includegraphics[width=0.45\textwidth]{lsps2a.epsi} 
\includegraphics[width=0.45\textwidth]{lsps2b.epsi} 
\caption{For the data in Col.~3 of Table~\ref{tab:hd154_magfield}, plus three previously published 
measurements of the magnetic field of HD\,154708 for the hydrogen lines, these plots  
show the standard deviation of one measurement with respect to a sinusoidal fit 
for a range of frequencies. The top panel shows that the best-fitting frequency is 
close to that determined in Fig.~\ref{fig:lsps1}, and the bottom panel shows at 
higher resolution that there is a best-fitting frequency with some alias 
ambiguities that fit less well. The best-fitting frequency is $f_{\rm rot} = 
0.18635 \pm 0.00008$\,d$^{-1}$, or $P_{\rm rot} = 5.367 \pm 0.002$\,d.
A weighted nonlinear least squares fit finally gives $P_{\rm rot} = 5.3666 \pm 0.0007$\,d
(see eq.~\ref{eq:5}) and discussion in the text.
} 
\label{fig:lsps2} 
\end{figure}

An initial frequency analysis was performed on the longitudinal magnetic field 
measurements given in Col.~2 of Table~\ref{tab:hd154_magfield} using the discrete Fourier transform 
programme of \citet{Kurtz1985}. The resulting amplitude spectrum clearly shows a 
dominant peak with an equivalent period near 5.3\,d, but the 13 measurements are 
not uniformly sampled over the rotation cycle for this period. We thus used the 
more appropriate frequency analysis technique of fitting a sinusoid by linear 
least-squares for a range of frequencies. Fig.~\ref{fig:lsps1} shows the 
unambiguous result. The best-fitting frequency of 0.18627\,d$^{-1}$ was then 
fitted by both linear least-squares and nonlinear least-squares sinusoidal fitting 
routines. The latter allows an estimate of the frequency uncertainty, and the 
former gives estimates of the amplitude and phase error of the fit. The resulting 
best fit gives the rotational frequency $f_{\rm rot} = 0.1863 \pm 0.0007$\,d$^{-
1}$, for a rotational period of $P_{\rm rot} = 5.367 \pm 0.020$\,d. The derived 
ephemeris for this period is:
\begin{equation}
\left<B_{\rm z}\right>^{\rm max} = {\rm MJD}54257.26 \pm 0.03 + 5.367 \pm 0.020 E
\end{equation}
To refine the period determination we combined the data
in Col.~3 of Table~\ref{tab:hd154_magfield} for 
the hydrogen lines with three previously published measurements made using the same 
technique obtained by \citet{Hubrig2006}. This data set has a time span of 3\,y, 
hence higher frequency resolution than the frequency analysis of the data obtained 
in a single season presented in Fig.~\ref{fig:lsps1}.  However, with only a few 
data points from previous years, the alias problem must be examined carefully. 
Because of the sparse distribution of the data, a frequency analysis was performed 
by fitting sinusoids by least-squares over an appropriate range of frequencies. As 
can be seen in Fig.~\ref{fig:lsps2}, there is some alias ambiguity, but it is not 
severe and a best-fitting frequency can be selected. As long as this is the 
correct alias, the rotation period is therefore refined to $P_{\rm rot} = 5.366 
\pm 0.002$\,d for HD\,154708. For this determination equal weights were used.
To increase the accuracy of the rotation period even further, we used a non-linear least-squares fit of 
the function $\overline{\left< B_{\rm z}\right>}+A_{\left< B_{\rm z}\right>} cos(2\pi(t/P+\phi))$ to 
the same 16~data points with weights equal to 
$1/\sigma^2$, with the $1\sigma$ uncertainties as listed in Col.~3 of Table~\ref{tab:hd154_magfield}.
We used the value $P=5.366$ found as the starting value for the fit 
and find $P_{\rm rot} = 5.3666 \pm 0.0007$ d.
We consider this rotation period as the most accurate. 
A similar fit was performed on the 13~data points obtained using the entire spectrum including 
all available absorption lines, but we kept in this case the period fixed.
The resulting final values are 
$\overline{\left< B_{\rm z}\right>} = 7199 \pm 11$, $A_{\left< B_{\rm z}\right>} = 789 \pm 13$ and $\phi=0.774 \pm 0.003$ 
with a reduced $\chi^2 = 3.6$.
This implies for the derived ephemeris for the maximum value of the magnetic field:

\begin{equation}
\left<B_{\rm z}\right>^{\rm max} = {\rm MJD}54257.24 \pm 0.02 + 5.3666 \pm 0.0007 E
\label{eq:5}
\end{equation}

\noindent
where the reference date is at the maximum closest to the middle of the observations presented in this paper. 

\begin{figure}
  \centering
  \includegraphics[width=0.45\textwidth]{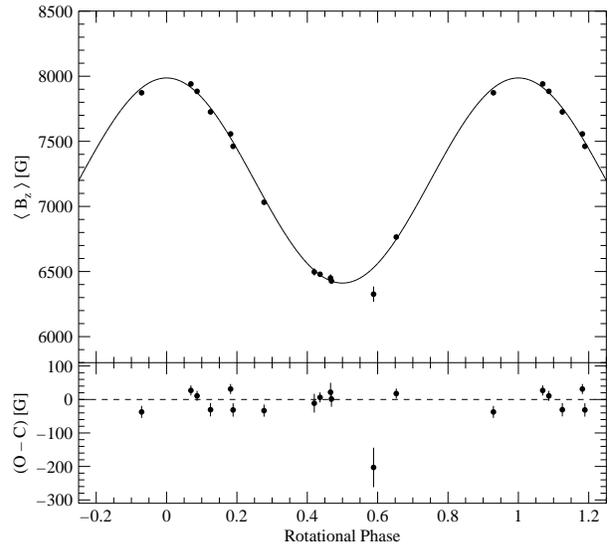}
  \caption{Phase diagram for the magnetic field measurements using 
all lines (Col.~2 of Table~\ref{tab:hd154_magfield})
   for the {\textbf most most accurate period, $P_{\rm rot} = 5.366$\,d.
   The variation has a mean of $\overline{\left< B_{\rm z}\right>} = 7199 \pm 11$\,G
   and an amplitude of $A_{\left< B_{\rm z}\right>} = 789 \pm 13$\,G.
The residuals (Observed -- Calculated) for the best fit is shown in the lower panel. The deviations are mostly of the 
same order as the error bars, and no systematic trends are obvious, which justifies a single sinusoid as a fit function. 
For the point near phase 0.6, see text.}}
  \label{fig:phase1}
\end{figure}

The corresponding phase diagram, including the best sinusoidal fit, is
shown in Fig.~\ref{fig:phase1}, along with the residuals (lower panel). Note that the size of 
the error bars
is comparable to the size of the dots representing the individual
measurements. The scatter of the latter about the best-fit curve is
also of the same order: this is a strong indication that the $1\sigma$
uncertainties given in Table~\ref{tab:hd154_magfield} are good
estimates of the actual random errors of the measurements
\citep{Landstreet1980}. To our knowledge, the present data
represent the first set of mean longitudinal magnetic field
measurements of a single star obtained with FORS\,1 with a sufficiently
uniform phase coverage to establish this important result. 
The measurement with the largest sigma 
($\left< B_{\rm z}\right>_{\rm all}=6326 \pm 59$\,G) was obtained in  weather conditions
classified as ``thick clouds'', where the guide star was frequently lost, and thus was 
repeated by the service observer a couple of nights later. Without taking into account 
this measurement we obtain a reduced $\chi^2 = 2.6$.

\section{Discussion} 
 
\begin{figure} 
\centering 
\includegraphics[width=0.45\textwidth]{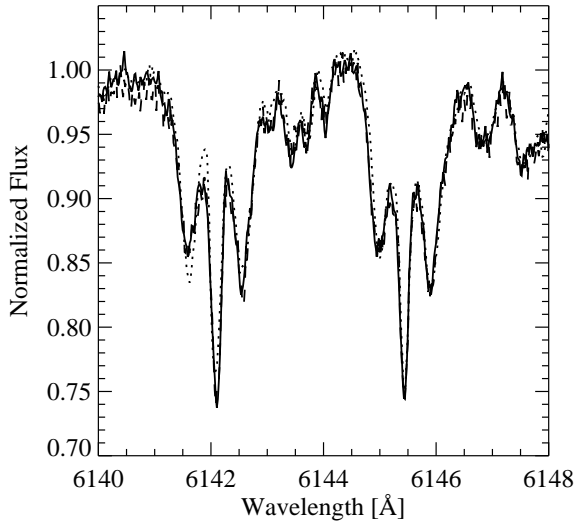} 
\caption{
One spectrum of HD\,154708 was obtained on May 20, 2005 at phase 0.72
and is indicated by the solid line.
The second spectrum was observed on September 18, 2005 at phase 0.37
and is presented by the dotted line.
The third spectrum was observed on July 23, 2006 at phase 0.76
and is presented by the dashed line.
No obvious changes are detectable in the split Zeeman structure
of the presented spectral lines.
}
\label{fig:Is} 
\end{figure} 
 
A rather short rotation period of 5.3666\,days is obtained for HD\,154708 based on 
the new thirteen low resolution spectropolarimetric observations. 
This is fully consistent with
the result found by \citet{Mathys1997} that mean magnetic field moduli
in excess of 7.5\,kG are only found in stars with rotation periods
shorter than 150 days. While many magnetic Ap stars have very
long rotation periods, up to several tens of years, or even close to 1
century in some cases, none of them have extremely strong fields such
as HD\,154708.

\citet{Hubrig2005} found  for HD\,154708 a very low value for \vsini{}. Our recent fit to 
the observed spectral line profile of the magnetically insensitive line 
Fe~I~$\lambda~5434.5$ resulted in $v \sin i  = 3.5 \pm 0.5$\,km\,s$^{-1}$.
This is consistent with the value of 4\,km\,s$^{-1}$ obtained by \citet{Elkin2008}.

Knowing the position of the star in the H-\nolinebreak{}R diagram \citep{Hubrig2005}, we 
determine the radius of the star to be in the range  $R=1.7-2.0$\,${\rm R}_\odot$. 
The equatorial rotation velocity is given by $v_{\rm eq} = 50.6 R/P$, where $R$ is 
the stellar radius in solar units and $P$ the period in days. From the measured $v 
\sin i$ values and $v_{\rm eq}$ determined using the known rotation period and the 
radius computed from the luminosity and effective temperature, the inclination of 
the stellar rotation axis of HD\,154708 can be determined. We obtain $v_{\rm 
eq}=16-18.9$\,km\,s$^{-1}$, leading to a value for the inclination angle in the range 
$i=9^{\circ}-14.5^{\circ}$. 

Our measurements exhibit a smooth, single-wave variation of the longitudinal 
magnetic field during the stellar rotation cycle. These observations can be 
considered as evidence for a dominant dipolar contribution to the magnetic field 
topology. Based on the assumption that the magnetic Ap stars are oblique dipole 
rotators, \citet{Preston1967} defined 
 
\begin{equation} 
r = \frac{\left< B_{\rm z}\right>^{\rm min}}{\left< B_{\rm z}\right>^{\rm max}} 
  = \frac{\cos \beta \cos i - \sin \beta \sin i}{\cos \beta \cos i + \sin \beta 
\sin i}, 
\end{equation} 
 
\noindent 
so that the obliquity angle, $\beta$ is given by
 
\begin{equation} 
\beta =  \arctan \left[ \left( \frac{1-r}{1+r} \right) \cot i \right]. 
\label{eqn:4} 
\end{equation} 
 
From the values determined above, $\left< B_{\rm z} \right>^{\rm max} = 7988 \pm 
17$\,G and $\left< B_{\rm z} \right>^{\rm min} = 6410 \pm 17$\,G, we find $r = 0.802 
\pm 0.004$, which for $i = 9^{\circ}-14.5^{\circ}$ leads to a magnetic obliquity 
in the range of $\beta= 22.5^{\circ}-35.5^{\circ}$.  
This value is close to the low end of the
range of obliquities determined by \citet{LandstreetMathys2000}
for magnetic Ap stars with rotation periods shorter than 25 days. 

For a tilted, centred magnetic dipole and assuming a limb darkening parameter of 
$u=0.5$, the surface polar field strength $B_{\rm d}$ is  
$B_{\rm d}\ge3.23\left<B_{\rm z}\right>^{\rm max}$
\citep{Preston1967}, i.e.\ $B_{\rm d}\ge$ 25.8\,kG 
for HD\,154708.
Using the values determined for $i$ and $\beta$, we find
$B_{\rm d}$=26.1\,kG for $i$=9$^{\circ}$ and $\beta$=35.5$^{\circ}$ and
$B_{\rm d}$=28.8\,kG for $i$=14.5$^{\circ}$ and $\beta$=22.5$^{\circ}$.
The measured  magnetic field modulus of $\sim$24.5\,kG 
using magnetically  resolved lines \citep{Hubrig2005} is of the same order as that 
estimated for the presented geometry of the magnetic field.

The limb-darkening coefficient is a function of wavelength.
Since no realistic model atmosphere
exists yet for the roAp stars with extremely strong magnetic fields, the
actual mean limb-darkening coefficient for the spectral range we
observed for HD\,154708 is uncertain. We note that the choice of other
mean limb-darkening coefficients would extend the range of possible
$B_{\rm d}$ values. 
The high spectral resolution UVES
spectra of HD\,154708 at three different dates at a resolving power  of
$\lambda{}/\Delta{}\lambda{} \approx 1.1\times10^5$
available in the ESO archive
show almost the same
appearance of magnetically resolved Zeeman components. In
Fig.~\ref{fig:Is} we present these three spectra 
where no significant changes in the split
line structures is visible, indicating no noticeable change in the
magnetic field modulus between these effectively two rotation phases
(since two of the spectra happen to be at nearly the same rotation
phase). The spectra obtained on May 20, 2005 at phase 0.72 and 
on July 23, 2006 at phase 0.76 are average spectra of the time
series used to study rapid oscillations in this star.
This does not represent a strong
constraint: in more than half of the known Ap stars with resolved
magnetically split lines, the ratio between the extrema of the mean
field modulus is lower than 1.1, and in 75\%\ of them, it does not
exceed 1.3. The three measurements of $\langle B\rangle$ that are
currently available are insufficient to define its variation, or to
check its consistency with the derived value of the rotation period.

\citet{LandstreetMathys2000} found a predominance of small values
of $\beta$ for stars with $P_{\rm rot} > 25$\,d. The only star with a
comparable field strength in their sample is Babcock's star HD\,215441,
which has $i$ and $\beta$ angles of the order of $\sim$30$^{\circ}$,
i.e.\ similar to the $i$ and $\beta$ determined for HD\,154708.
Babcock's star is much hotter and more massive than HD\,154708, and its
rotation period is somewhat longer at ($P_{\rm rot} = 9.5$\,d) than that
of HD\,154708. It is premature to make any statistical assertion about
the efficiency of magnetic braking and field geometry evolution just
from the knowledge of these two strongly magnetic stars. On the other
hand, \citet{Hubrig2007} studied the behaviour of the angle $\beta$ in
Ap stars of different masses and ages. The obliquity angle distribution
as inferred from the distribution of $r$-values appears random at the
time magnetic stars become observable on the H-R diagram. After only a
short time spent on the main sequence, the obliquity angle $\beta$ then
tends to reach values close to either 90$^\circ$ or 0$^\circ$ for ${\rm
M}<3$\,${\rm M}_\odot$.


As we already mentioned above, due to the presence of the very strong magnetic field 
in the atmosphere of HD\,154708, its spectrum is very complex and many spectral 
lines need a detailed  Paschen-Back treatment. On the other hand, due to the low value of 
\vsini{} and the relatively short rotation  period, this star is one of the 
most suitable targets to study various atmospheric effects that interact with a 
strong magnetic field.
 

 
\label{lastpage} 
 
\end{document}